\documentclass[prd,twocolumn,aps,amsmath,nofootinbib,superscriptaddress]{revtex4}
\usepackage{graphicx}
\usepackage{bm}
\usepackage{epsfig}

  \newcount\hour \newcount\minute
  \hour=\time \divide \hour by 60
  \minute=\time
  \newcommand{\mydate}{\ \today \ - \number\hour :\ifnum \minute<10 0\fi 
\number\minute}

\def\nslash{n\!\!\!\slash}
\def\bnslash{\bar n\!\!\!\slash}

\def\OMIT#1{}

\newcommand{\nn}{\nonumber} 

\newcommand{\bn}{{\bar n}}
\newcommand{\bea}{\begin{eqnarray}}
\newcommand{\eea}{\end{eqnarray}}

\def\lqcd{\Lambda_{\rm QCD}}

\begin{document}



\title{Graphical amplitudes from SCET}

\author{Christian W.~Bauer}
\affiliation{California Institute of Technology, Pasadena, CA 91125}
\author{Dan Pirjol}
\affiliation{Center for Theoretical Physics, Massachusetts Institute of
  Technology, Cambridge, MA 02139}

\begin{abstract}

We discuss the relationship between the graphical amplitudes 
$T$, $C$, $P\, \ldots$ used to parameterize nonleptonic $B$ decay 
amplitudes,  and matrix elements
of operators in the soft-collinear effective theory (SCET) at leading order in 
$\Lambda/m_b$. Using the SU(3) flavor symmetry of the weak Hamiltonian we derive 
all-order constraints on the Wilson coefficients of SCET operators.

\end{abstract}

\maketitle

There are several theoretical approaches to treating exclusive non-leptonic $B$ decays to two
or more mesons. A first class of approaches uses flavor symmetries of QCD, in particular 
SU(3)~\cite{Zeppenfeld:1980ex,Savage:1989ub}. 
One then decomposes all 
possible amplitudes into the general set of reduced matrix elements in SU(3), and all relative
factors can then be obtained from group theory. An alternative approach is the use of so called
graphical amplitudes, which are defined by the topology of the quarks in a given 
diagram~\cite{Chau:1990ay,Gronau:1994rj,Gronau:1998fn}. It was shown that there 
exists an equivalence between the graphical amplitudes method and the SU(3) reduced
matrix elements~\cite{Zeppenfeld:1980ex}, however a complete classification of all
possible Wick contractions is rather complex \cite{BuSi}. The main complication is due to
the appearance of rescattering effects 
\cite{rescatt}. For example, the color-suppressed decay
$\bar B^0\to D^0\pi^0$ can proceed through an intermediate color-allowed hadronic channel
$\bar B^0\to D^+\pi^-$, followed by strong rescattering. In many analyses dynamical assumption 
about the sizes of the various graphical amplitudes are made, and certain graphical amplitudes
are omitted.

A second class of approaches goes beyond using flavor symmetries and uses
the limit $m_b\to \infty$ to simplify the amplitudes. This was started with the perturbative 
QCD~\cite{Keum:2000wi} method and the QCD factorization approach~\cite{Beneke:1999br}, and 
recently an effective theory treatment of these decays in the framework of soft 
collinear effective theory (SCET)~\cite{SCET} was developed~\cite{Chay:2003zp,Bauer:2004tj}. 
In this letter we show that there is a simple relationship between the graphical amplitudes 
and matrix elements of SCET operators, and we will prove the dynamical assumptions usually made. 
The use of isospin and SU(3) symmetry in SCET is also discussed.

The electroweak Hamiltonian mediating non-leptonic $B$ decays is given by the
$\Delta B=1$ Hamiltonian, which reads
\begin{eqnarray} \label{HW}
 H_W = \frac{G_F}{\sqrt{2}} \left[ \sum_{p=u,c} \lambda_p
 \left( C_1 O_1^p + C_2 O_2^p  \right)
  - \lambda_t\sum_{i=3}^{10}\!\! C_i O_i \Big) \right],\nn
\end{eqnarray}
where the CKM factor is $\lambda_p = V_{pb} V^*_{pf}$ with $f=d,s$ for $\Delta S = 0,1$
transitions, respectively.  The operators $O_{1,2}$ are the well known current-current operators
\begin{eqnarray}\label{O12}
 O_1^p \!\! &=&\!\! (\overline{p} b)_{V\!-\!A}
  (\overline{f} p)_{V\!-\!A}, \ \
 O_2^p = (\overline{p}_{\beta} b_{\alpha})_{V\!-\!A}
  (\overline{f}_{\alpha} p_{\beta})_{V\!-\!A}, \nonumber 
\end{eqnarray}
the operators 
$O_{3-6}$ are known as strong penguin operators
\begin{eqnarray}\label{O34}
O_{3,4} \!\! &=& \!\! \big\{ (\overline{f} b)_{V\!-\!A}
  (\overline{q} q)_{V\! - \!A}\,, (\overline{f}_{\beta} b_{\alpha})_{V\!-\!A}
  (\overline{q}_{\alpha} q_{\beta})_{V\! - \!A} \big\}, \nonumber  \\
 O_{5,6} \!\! &=& \!\! \big\{ (\overline{f} b)_{V\!-\!A}
  (\overline{q} q)_{V\! + \!A}\,, (\overline{f}_{\beta} b_{\alpha})_{V\!-\!A}
  (\overline{q}_{\alpha} q_{\beta})_{V\! + \!A} \big\}, \nonumber 
\end{eqnarray}
and the remaining operators denote electroweak penguin operators, whose definition
can be obtained from Ref.~\cite{Buchalla:1995vs}. 

The graphical amplitudes are defined as follows~\cite{Chau:1990ay,Gronau:1994rj,Gronau:1998fn}
\begin{itemize}
\item {\it Tree amplitude $T$:}\\
Operators $O_{1,2}^u$ with the $f$ quark and the $\bar u$ quark of operator forming a meson.\\[-.4cm]
\item {\it Color suppressed amplitude $C$:}\\
Operators $O_{1,2}^u$ with the $f$ quark and the spectator quark of the $B$ forming a meson. 
\item {\it Penguin amplitudes $P^{u}$, $P^c$, or $P^t$:}\\
Operators $O_{1,2}^u$, $O_{1,2}^c$  with the two identical light quarks forming a
virtual loop, or contributions from $O_{3-6}$, where in all three cases the $f$ quark ends up in 
one of the light mesons.
\item {\it Annihilation amplitude $A$:}\\
Operators $O_{1,2}^u$ with the $u$ and $\bar b$ quark of operator forming the $B$ meson.
\item {\it Exchange amplitude $E$:}\\
Operators $O_{1,2}^u$ with the $f$ and $\bar b$ quark of operator forming the $B$ meson.
\item {\it Penguin annihilation amplitude $PA^{u,c,t}$:}\\
Operators $O_{1,2}^u$, $O_{1,2}^c$  with the two identical light quarks forming a
virtual loop or contributions from $O_{3-6}$, with the $f$ quark ending up in 
the $B$ mesons.
\item {\it Electroweak penguin amplitudes $EW^{T}$:}\\
Operators $O_{7-10}$ with the $f$ quark and the spectator quark of the $B$ forming a meson.
\item {\it Electroweak penguin amplitudes $EW^{C}$:}\\
Operators $O_{7-10}$ with the $f$ quark and the $\bar q$ quark forming a meson.
\item {\it Electroweak penguin amplitudes $EW ^{P}$:}\\
Operators $O_{7-10}$ with the $f$ quark and the $\bar q$ quark forming a virtual loop. 
\item {\it Electroweak penguin amplitudes $EW ^{A,E,PA}$:}\\
Operators $O_{7-10}$ with the topologies identical to the $A,E,PA$ topologies defined above.
\end{itemize}
All amplitudes in $B \to M_1 M_2$ decays (with $M_i$ being any light 
meson) can be written in terms of these graphical amplitudes. These relations are given
in tabular form in Refs.~\cite{Gronau:1994rj,Gronau:1998fn}. The SU(3) analysis in 
terms of graphical amplitudes is often supplemented with dynamical
assumptions about their sizes, and some graphical amplitudes are typically neglected.

In SCET one uses the fact that the two light mesons have large energy much larger than $\lqcd$.
This allows to match the Hamiltonian given in Eq.~(\ref{HW}) at the scale $m_b$ onto the effective
theory Hamiltonian 
\begin{eqnarray} \label{match}
 H_W \!\!&=  &\!\! \frac{2G_F}{\sqrt{2}} \sum _{n,\bn} \bigg\{ 
  \sum_i \int [d\omega_{j}]_{j=1}^{3}
       c_i(\omega_j)  Q_{i}^{(0)}(\omega_j) \nn\\ 
 && \hspace{-1cm}
  + \sum_i \int [d\omega_{j}]_{j=1}^{4}  b_i(\omega_j) 
  Q_{i}^{(1)}(\omega_j) 
  + {\cal Q}_{c\bar c} + \ldots \bigg\} \,,
\end{eqnarray}
where $c_i^{(f)}(\omega_j)$ and $b_i^{(f)}(\omega_j)$ are Wilson coefficients, the ellipses
denote color-octet operators which do not contribute at leading order and higher order terms 
in $\lqcd/Q$, $Q=\{m_b,E\}$, and ${\cal Q}_{c\bar c}$
denotes operators containing a $c \bar c$ pair. In the remainder of this paper we often
omit the dependence of the Wilson coefficients on the $\omega_j$. 
In Eq.~(\ref{match}) the ${\cal O}(\lambda^0)$ operators are [sum over
$q=u,d,s$]
\begin{eqnarray} \label{Q0}
  Q_{1}^{(0)} &=&  \big[ \bar u_{n,\omega_1} \bnslash P_L b_v\big]
  \big[ \bar f_{\bn,\omega_2}  \nslash P_L u_{\bn,\omega_3} \big]
  \,,  \\
  Q_{2,3}^{(0)} &=&  \big[ \bar f_{n,\omega_1} \bnslash P_L b_v \big]
  \big[ \bar u_{\bn,\omega_2} \nslash P_{L,R} u_{\bn,\omega_3} \big]
   \,,\nn \\
  Q_{4}^{(0)} &=&  \big[ \bar q_{n,\omega_1} \bnslash P_L b_v \big]
  \big[ \bar f_{\bn,\omega_2} \nslash P_{L}\, q_{\bn,\omega_3} \big]
   \,, \nn 
\end{eqnarray}
where we have omitted operators which give rise to flavor-singlet light mesons. Note that the
operator $Q_3$ is only required if electroweak penguins are included. 
The effective theory operators contain collinear fields along both $n$ and $\bn$
directions \cite{nbn}.
Also required are operators suppressed by one 
power of the SCET expansion parameter, as explained in~\cite{Bauer:2004tj}. They are given by
\begin{eqnarray}\label{Q1a1b}
  Q_{1}^{(1)} \!\!&=&\!\! \frac{-2}{m_b} 
     \big[ \bar u_{n,\omega_1}\, ig\,\slash\!\!\!\!{\cal B}^\perp_{n,\omega_4} 
     P_L b_v\big]
     \big[ \bar f_{\bn,\omega_2}  \nslash P_L u_{\bn,\omega_3} \big] 
     \,, \\
  Q_{2,3}^{(1)} &=&  \frac{-2}{m_b}  
     \big[ \bar f_{n,\omega_1} \, ig\,\slash\!\!\!\!{\cal B}^\perp_{n,\omega_4} 
     P_L b_v \big]
     \big[ \bar u_{\bn,\omega_2} \nslash P_{L,R} u_{\bn,\omega_3} \big]
      \,,\nn \\
  Q_{4}^{(1)} &=&  \frac{-2}{m_b} 
     \big[ \bar q_{n,\omega_1} \, ig\,\slash\!\!\!\!{\cal B}^\perp_{n,\omega_4} 
     P_L b_v \big]
     \big[ \bar f_{\bn,\omega_2} \nslash P_{L}\, q_{\bn,\omega_3} \big]
      \,.\nn 
\end{eqnarray}

The factorization properties of SCET can be used
to simplify the matrix elements of these operators significantly, leading to a final expression
for the amplitude of an arbitrary $B \to M_1 M_2$ decay~\cite{Bauer:2004tj}
\begin{eqnarray}  \label{A0newfact}
A \!\!&=&\!\! 
N_0\bigg\{
   f_{M_1}\! \int_0^1\!\!\!\!du\, dz\,
    T_{1\!J}(u,z) \zeta^{BM_2}_{J}(z) \phi^{M_1}(u) 
   \\
 &&\hspace{-0.7cm}
   + f_{M_1} \zeta^{BM_2}\!\! \int_0^1\!\!\!\! du\, T_{1\zeta}(u) \phi^{M_1}(u)
  \bigg\} \!+\! \Big\{ 1\leftrightarrow 2\Big\} 
 \!+\! \lambda_c^{(f)} A_{c\bar c}^{M_1M_2}. \nn 
\end{eqnarray}
Here $A_{c\bar c}^{M_1M_2} = \langle M_1 M_2 | {\cal Q}_{c \bar c} | B \rangle$, $N_0 = \frac{G_F m_B^2}{\sqrt2}$,
the $f_{M}$'s are decay constants of the meson $M$ and the  $\zeta^{BM}$ and 
$\zeta^{BM}_J(z)$ are transition matrix elements between the initial $B$ meson and a final 
meson $M$. They are of the same order in the power counting and are identical to
the non-perturbative parameters which appear in $B \to M$ form factors \cite{Bauer:2002aj}. 

The Wilson coefficients of the operators $Q_{1-4}^{(0,1)}$ are given at tree level in~\cite{Bauer:2004tj}. To 
separate the contributions of full theory current-current operators, QCD penguin operators and 
electroweak penguin operators to 
the SCET matching coefficients, it
is convenient to write the Wilson coefficients of the SCET operators as
\begin{eqnarray}
c_i &=& \lambda_u c_{iu} + 
\lambda_t \left[c_{it}^{\rm p} + c_{it}^{\rm ew} \right] \nn\\
b_i &=& \lambda_u c_{iu} + 
\lambda_t^{(f)} \left[b_{it}^{\rm p} + b_{it}^{\rm ew} \right] \,.
\end{eqnarray}
Here $c_{it}^{\rm p}$, $b_{it}^{\rm p}$ denote the terms 
proportional to the full theory Wilson coefficients $C_{3-6}$ of the QCD penguin operators, 
while  $c_{it}^{\rm ew}$,$b_{it}^{\rm ew}$ denote the terms 
proportional to the full theory Wilson coefficients $C_{7-10}$ of the electroweak penguin operators.
With these definitions, the tree level Wilson coefficients are
\begin{eqnarray}\label{cexpr}
 &&c_{1u}=
      C_1 \!+\! \frac{C_2}{N_c}\,,\qquad
 c_{1t}^{\rm ew} =
       - \frac32 \Big( C_{10} \!+\! \frac{C_9}{N_c}\Big)\nn\\
 &&c_{2u} =
      C_2 \!+\! \frac{C_1}{N_c}\,,\qquad
 c_{2t}^{\rm ew}= - \frac32 \Big(C_9 \!+\! \frac{C_{10}}{N_c}\Big)
      \nn \\
 &&
 c_{3t}^{\rm ew} =
       - \frac32  \Big( C_7 + \frac{C_8}{N_c}\Big)\nn\\
 &&
 c_{4t}^p = - \Big(C_4 + \frac{C_3}{N_c}\Big)\,,\qquad
 c_{4t}^{\rm ew} =\frac{1}{2} \Big(C_{10} + \frac{C_9}{N_c}\Big) \,,\quad\nn\\
 &&
 c_{1t}^p = c_{2t}^p = c_{3u} =c_{3t}^p = c_{4u}= 0
\end{eqnarray}
and 
\begin{eqnarray}\label{bexpr}
 &&b_{1u}=
      C_1 + \Big(1 \!-\!\frac{m_b}{\omega_3} \Big)
      \frac{C_2}{N_c}\nn\\
 &&
 b_{1t}^{\rm ew} =
       - \frac32 \Big[C_{10} + 
    \Big(1 \!-\!\frac{m_b}{\omega_3} \Big)
      \frac{C_9}{N_c} \Big]\nn\\
 &&b_{2u} =
      C_2 + \Big(1 \!- \! \frac{m_b}{\omega_3}  \Big)
      \frac{C_1}{N_c}\nn\\
&&
b_{2t}^{\rm ew}= - \frac32 \Big[ C_9 + \Big(1 \!
     -\!\frac{m_b}{\omega_3} \Big)
      \frac{C_{10}}{N_c} \Big]
      \nn \\
 &&
 b_{3t}^{\rm ew} =
       - \frac32  \Big[ C_7 
     + \Big(1 \!-\!\frac{m_b}{\omega_2} \Big)
      \frac{C_8}{N_c} \Big]\nn\\
 &&
 b_{4t}^p = - C_4 - \Big(1 \!-\!\frac{m_b}{\omega_3} \Big)\frac{C_3}{N_c}\nn\\
 &&
 b_{4t}^{\rm ew} =\frac{1}{2} \Big[C_{10} + \Big(1 \!-\!\frac{m_b}{\omega_3} \Big)\frac{C_9}{N_c} \Big]\,.\quad\nn\\
&&
b_{1t}^p = b_{2t}^p = b_{3u} =b_{3t}^p = b_{4u}= 0
\end{eqnarray}
The Wilson coefficients $c_{1t}^p, c_{2t}^p$ and
$b_{1t}^p, b_{2t}^p$ vanish to all orders in matching, since
the penguin operators $O_{3-6}$ transform as a $\mathbf{3}$ 
under flavor $SU(3)$, while $Q_{1,2}$ transform as a combination
of $\mathbf{3,\overline{6},15}$. A similar argument gives $c_{4u}=b_{4u}=0$.
Chirality invariance requires that $c_{3u}=b_{3u}=0$ to all orders.

The graphical amplitudes are defined by the contractions of the quark lines of the operator with
the quarks in the mesons. In QCD, this separation is difficult to define in an unambiguous way 
in terms of matrix elements of operators. SCET however allows this distinction, since each 
meson consists of interpolating fields with well defined collinear directions. The $B$ meson is described by an 
interpolating field of two soft quarks, while the two light mesons are defined by interpolating 
fields of two collinear quarks in different light-like directions. At leading order in SCET there
are no couplings between soft and collinear particles, or between collinear particles in different directions. 
Since the four quarks in the SCET operators in Eqs.~(\ref{Q0}) and~(\ref{Q1a1b}) are either soft
or have a well 
defined collinear direction, each of them ends up in a particular meson. Thus, the 
graphical amplitudes can be defined as matrix elements of SCET operators. 

The relation of the graphical amplitudes to SCET matrix elements can be obtained
without making use of any flavor symmetries.  While this will result in too many graphical 
amplitudes to have any predictive power, it will illustrate the identification between graphical
amplitudes and SCET matrix elements. We will later show how flavor symmetries can be 
used to reduce the set of amplitudes. We add subscripts $M_1M_2$ to any graphical
amplitude to denote the final state of the decay. A general amplitude defined as 
$A_{M_1 M_2} \equiv A(B \to M_1 M_2)$ is then given by
\begin{eqnarray}
A_{M_1 M_2} &=& \lambda_u (T_{M_1M_2} + C_{M_1M_2} +A_{M_1M_2} + E_{M_1M_2} 
\nn\\
&&
\qquad \qquad+ P^u_{M_1M_2} + PA^u_{M_1M_2} ) 
\nn\\
&&
+ \lambda_c (P^c_{M_1M_2}+PA^c_{M_1M_2}) 
\nn\\
&&
+ \lambda_t (P^t_{M_1M_2}+ PA^t_{M_1M_2} + EW_{M_1M_2} )\,.
\end{eqnarray}
The graphical amplitudes are defined as before, but depend on the particular final state. Here $EW$
denotes the sum of all possible electroweak penguin amplitudes $EW^{T,C,P,A,E,P\!A}$. 

So far the matching contributions from charming penguins have not been fully worked out in SCET, 
thus their effects are not included in the SCET operators $Q_{1-4}^{(0,1)}$. Thus, we make 
no attempt here to derive explicit results for $P^c_{M_1M_2}$ and $PA^c_{M_1M_2}$ and 
simply leave them as unknown parameters
\begin{eqnarray}
P^c_{M_1M_2} = A^{\rm P}_{cc_{M_1M_2}}\,,\qquad 
P\!A^c_{M_1M_2} = A^{\rm PA}_{cc_{M_1M_2}}\,.
\end{eqnarray}
In $B\to \pi\pi$ decays the charming penguins appear always in the
combination $ A^{\rm P}_{cc_{\pi\pi}}  + A^{\rm PA}_{cc_{\pi\pi}} = A_{cc}^{\pi\pi}$ 
where $A_{cc}^{\pi\pi}$ is the amplitude introduced in Ref.~\cite{Bauer:2004tj}. 
However, in $B\to K\pi$ only the $ A^{\rm P}_{cc_{K\pi}}$ amplitude
contributes. We will neglect here charming
penguin effects arising from electroweak penguins due to their small Wilson coefficients.

Next, consider the tree amplitudes $T$ and $C$. Since they originate
from the current-current operators, and do not contain a penguin contraction, 
the operators $Q^{(0,1)}_{4}$ do not contribute. Furthermore, since the 
current-current operators are 
multiplied by $\lambda_u$, only the Wilson coefficients $c_{1u-3u}$  and 
$b_{1u-3u}$ are required. Finally, since the operator $Q_3$ only receives 
contributions from electroweak penguin operators
in the full theory, $c_{3u} = b_{3u} = 0$. 
The distinction between the $T$ and $C$ amplitudes is whether the quark $f$ forms a 
light meson with a $\bar u$ quark of the operator or with the spectator of the $B$ meson. Since the 
two collinear quarks in the $\bn$ direction form one meson, while the collinear quark in the $n$ 
direction combines with the spectator of the $B$ in the second step of matching, the contributions
to $T$ are only from operators $Q^{(0,1)}_{1}$ and those to $C$ are from $Q^{(0,1)}_{2}$.
This gives
\begin{eqnarray}
T_{M_1M_2} &=& N_0   \langle M_1M_2 | c_{1u} Q^{(0)}_1 
+ b_{1u} Q^{(1)}_1 | B \rangle \nn\\
C_{M_1M_2} &=& N_0  \langle M_1M_2 | c_{2u} Q^{(0)}_2 
+ b_{2u} Q^{(1)}_2 | B \rangle \,.
\end{eqnarray}
In writing the matrix element of $Q^{(0,1)}_i$ it is understood that the definition
of the matrix elements involves time-ordered products with subleading interactions in SCET which 
couple a soft to a collinear quark. For details see~\cite{Bauer:2002aj,Bauer:2004tj}.

\begin{table}[t]
\begin{tabular}{c}
$B \to \pi \pi$\\
\hline\hline
\begin{tabular}{c||c|c|c|c||ccc}
mode & $T_{\pi\pi}$ & $C_{\pi\pi}$ & $P^{u,c,t}_{\pi\pi}$ & $P\!A^c_{\pi\pi}$ &
$EW^T_{\pi\pi}$ & $EW^C_{\pi\pi}$ 
& $EW^P_{\pi\pi}$ \\
\hline
$\pi^+ \pi^-$ & $-1$ & 0 & $-1$ & $-1$ & $0$ & $1$ & $-1$\\
$\pi^+ \pi^0$ & $-\frac{1}{\sqrt{2}} $ & $-\frac{1}{\sqrt{2}} $ & 0 & 0 & $\frac{1}{\sqrt{2}} $ 
& $\frac{3}{2\sqrt{2}} $ & $0$\\
$\pi^0 \pi^0$ & 0 & $-1$ & $1$ & $1$ & $1$ & $\frac12$ & $1$ \\
\hline\hline
\end{tabular}
\end{tabular}
\caption{SU(2) relations for the decays $B \to \pi \pi$
\label{tab:pipi}}
\begin{tabular}{c}
$B \to K \pi$\\
\hline\hline
\begin{tabular}{c||c|c|c|c||ccc}
mode & $T_{K\pi}$ & $C_{K\pi}$ & $P^{u,c,t}_{K\pi}$ & $P\!A^c_{K\pi}$ &
$EW^T_{K\pi}$ & $EW^C_{K\pi}$ & $EW^P_{K\pi}$ \\
\hline
$K^- \pi^+$ & $-1$ & 0 & $-1$ & 0 & $0$ & $1$  & $-1$\\
$K^- \pi^0$ & $-\frac{1}{\sqrt{2}} $ & $-\frac{1}{\sqrt{2}} $ & $-\frac{1}{\sqrt{2}} $ & 0 
& $\frac{1}{\sqrt{2}}$ & $\frac{1}{\sqrt{2}}$  & $-\frac{1}{\sqrt{2}} $\\
$\bar K^0 \pi^-$ & 0 & $0$ & $1$ & 0 & $0$ & $\frac12$  & $1$\\
$\bar K^0 \pi^0$ & 0 & $-\frac{1}{\sqrt{2}}$ & $\frac{1}{\sqrt{2}}$ & 0 
& $\frac{1}{\sqrt{2}}$ & $\frac{1}{2\sqrt{2}}$  & $\frac{1}{\sqrt{2}}$ \\
\hline\hline
\end{tabular}
\end{tabular}
\caption{SU(2) relations for the decays $B \to K \pi$
\label{tab:Kpi}}
\begin{tabular}{c}
$B \to K \bar K$\\
\hline\hline
\begin{tabular}{c||c|c|c|c||ccc}
mode & $T_{K\bar K}$ & $C_{K\bar K}$ & $P^{u,c,t}_{K\bar K}$ & 
$P\!A^{c}_{K\bar K}$ & $EW^T_{K\bar K}$ & $EW^C_{K\bar K}$ & $EW^P_{K\bar K}$ \\
\hline
$K^+ K^-$ & $0$ & 0 & $0$ & $-1$ & $0$ & 0 & 0\\
$K^- K^0$ & $0$ & $0$ & 1 & 0 & $0$ & $\frac12$ & 1\\
$\bar K^0 K^0$ & 0 & $0$ & $1$ & 1 & $0$ & $\frac12$ & 1 \\
\hline\hline
\end{tabular}\\
\end{tabular}
\caption{SU(2) relations for the decays $B \to K \bar K $
\label{tab:KK}}
\end{table}

The other graphical amplitudes can be obtained in the same way. Since the three light quarks
of the operators in SCET are all collinear, they can not be part of the $B$ meson. Thus, 
to leading order in SCET we find
\begin{eqnarray}
A_{M_1M_2} = E_{M_1M_2} = P\!A^{u,t}_{M_1M_2} = 0\,.
\end{eqnarray}
The QCD penguin contribution are from matrix elements of the operator $Q_4^{(0,1)}$ and we find
\begin{eqnarray}
P^u_{M_1M_2} &=& N_0   \langle M_1M_2 | c_{4u} Q^{(0)}_4 
+ b_{4u} Q^{(1)}_4 | B \rangle \nn\\
P^t_{M_1M_2} &=& N_0   \langle M_1M_2 | c_{4t}^{\rm p} Q^{(0)}_4 
+ b_{4t}^{\rm p} Q^{(1)}_4 | B \rangle \,.
\end{eqnarray}

Finally, we need the electroweak penguin amplitudes. 
We find 
\begin{eqnarray}
EW^T_{M_1M_2} &=& N_0   \langle M_1M_2 |
\sum_{j=2,3} c_{jt}^{\rm ew} Q^{(0)}_j
+ b_{jt}^{\rm ew} Q^{(1)}_j | B \rangle \nn\\
EW^C_{M_1M_2} &=& N_0  \langle M_1M_2 |(c_{1t}^{\rm ew}/3) ( 3Q^{(0)}_1 - Q^{(0)}_4)  \nn\\
&& \qquad \qquad +  (b_{1t}^{\rm ew}/3) ( 3Q^{(1)}_1 - Q^{(1)}_4) | B \rangle \nn\\
EW^P_{M_1M_2} &=& N_0  \langle M_1M_2 | (c_{4t}^{\rm ew}+c_{1t}^{\rm ew}/3)  Q^{(0)}_4 
\nn\\
&&\qquad \qquad
+ 
(b_{4t}^{\rm ew} + b_{1t}^{\rm ew}/3) Q^{(1)}_4
 | B \rangle \,,
\end{eqnarray}
and at leading order
\begin{eqnarray}
EW^A_{M_1M_2} \!= EW^E_{M_1M_2} \!= 
EW^{PA}_{M_1M_2} \!= 0\,.
\end{eqnarray}

So far our results for the graphical amplitudes have been general, and no symmetry 
assumption has been made.
Since the number of graphical amplitudes is larger than the number of decay 
modes, there is no predictive power. We
start by assuming isospin symmetry. This gives a reduction in the 
number of amplitudes, which are labeled by the isospin content of the
final state $M_1 M_2$. For example, all modes with two pions in the final state 
have graphical amplitudes labeled by the isospin content $_{\pi\pi}$, 
for $K\pi$ final states we label them by $_{K \pi}$, and so on. In Tables~\ref{tab:pipi}-\ref{tab:KK} 
we give the relative
factors between the graphical amplitudes for a given final state to the graphical 
amplitudes with given isospin content. 
These tables agree with the results in Ref.~\cite{Gronau:1994rj}, 
if one takes into account that the EWP graphical amplitudes 
$EW^{T,C}$ are defined with a different normalization than in
Ref.~\cite{Gronau:1994rj}. We use the same definitions for $PA^{c}$
as in Refs.~\cite{Gronau:1994rj} and reproduce here the results for 
these amplitudes for completeness.
The results for final states with two vector 
mesons are identical to those for two pseudoscalar mesons. 

The amplitudes of given isospin content can be expressed 
in terms of SCET parameters. These relations have a similar form
for $PP$ and $VV$ final states. For flavor content $K\pi$ they are
given by
\begin{eqnarray}\label{GPP}
T_{K\pi} &\!\!\!=\!\!\!& -N_0 
 f_{K} \left[ \langle c_{1u} \rangle_K 
\zeta^{B\pi} + \langle b_{1u}\zeta_J^{B\pi}\rangle_{K}  \right]\\
C_{K\pi} &\!\!\!=\!\!\!& -N_0 
 f_\pi \left[ \langle c_{2u}\rangle_\pi 
\zeta^{BK} + \langle b_{2u} \zeta_J^{BK} \rangle_\pi  \right]\nn\\
P^{u}_{K\pi} &\!\!\!=\!\!\!&  -N_0 f_K
\left[ \langle c_{4u}\rangle_K 
\zeta^{B\pi} + \langle b_{4u}  \zeta_J^{B\pi} \rangle_K\right]\nn\\
P^{t}_{K\pi} &\!\!\!=\!\!\!&  -N_0 f_K
\left[ \langle c_{4t}^{\rm p}\rangle_K 
\zeta^{B\pi} + \langle b_{4t}^{\rm p} \zeta_J^{B\pi} \rangle_K \right]\nn\\
EW^T_{K\pi} &\!\!\!=\!\!\!& N_0
f_\pi \left[ \langle c_{2t}^{\rm ew} - c_{3t}^{\rm ew} \rangle_\pi
\zeta^{BK} + \langle (b_{2t}^{\rm ew} - b_{3t}^{\rm ew} )
 \zeta_J^{BK} \rangle_\pi\right]\nn\\
EW^C_{K\pi} &\!\!\!=\!\!\!& \frac{2N_0}{3}
f_K \left[ \langle c_{1t}^{\rm ew} \rangle_K
\zeta^{B\pi} + \langle b_{1t}^{\rm ew} \zeta_J^{B\pi} \rangle_K \right]\nn\\
EW^P_{K\pi} &\!\!\!=\!\!\!& -\frac{N_0}{3}
f_K \left[ \langle 3\,c_{4t}^{\rm ew}+ c_{1t}^{\rm ew} \rangle_K
\zeta^{B\pi} 
\right.\nn\\
&&\qquad \qquad \qquad \left.
+ \langle (3 \,b_{4t}^{\rm ew} + b_{1t}^{\rm ew}) \zeta_J^{B\pi} \rangle_K \right]\,.\nn
\end{eqnarray}
The results for $K^* \rho$ are the same with changed subscripts for the flavor content 
and an opposite sign between the two Wilson coefficients [$(c_{2t}^{\rm ew} + c_{3t}^{\rm ew})$,  $(b_{2t}^{\rm ew} + b_{3t}^{\rm ew})$] in $EW^T_{K\pi}$.  The other channels
$\pi\pi, K\bar K,\rho\rho,K^*\bar K^*$
can be obtained from this by straightforward substitutions. 
For simplicity of notation we defined
\begin{eqnarray}
\langle c_{iu}\rangle_{M_1} &=& \int_0^1 du c_{iu}(u) \phi_{M_1}(u) \,,\\
\langle b_{iu}\zeta_J^{BM_2} \rangle_{M_1} &=& \int_0^1 du dz b_{iu}(u,z) \phi_{M_1}(u) 
\zeta_J^{BM_2}(z)\,.\nn
\end{eqnarray}

\begin{table*}[t]
\begin{tabular}{c}
$\bar B \to \rho \pi$ \\
\hline\hline
\begin{tabular}{c|c|c|c|c|c|c|c|c||cccccc}
mode & $T^P_{\rho\pi}$ & $T^V_{\rho\pi}$ & $C^P_{\rho\pi}$ & 
$C^V_{\rho\pi}$ & $P^{(u,c,t)P}_{\rho\pi}$ & $P^{(u,c,t)V}_{\rho\pi}$ & 
$PA^{cP}_{\rho\pi}$ & $PA^{cV}_{\rho\pi}$ & 
$EW^{TP}_{\rho\pi}$ &  $EW^{TV}_{\rho\pi}$ & 
$EW^{CP}_{\rho\pi}$ &  $EW^{CV}_{\rho\pi}$ & 
$EW^{PP}_{\rho\pi}$ &  $EW^{PV}_{\rho\pi}$ \\
\hline
$\rho^+ \pi^-$ & 0 & $-1$ & 0 & 0 & 0 & $-1$ & $-1$ & $-1$ & 0 & 0 & 0 & 1 & 0 & $-1$\\
$\pi^+ \rho^-$ & $-1$ & 0 & 0 & 0 & $-1$ & 0 & $-1$ & $-1$ & 0 & 0 & 1 & 0 & $-1$ & 0\\
$\rho^0 \pi^0$ & 0 & 0 & $-\frac12$ & $-\frac12$ & $\frac12$ & $\frac12$ & $\frac12$ & 
$\frac12$ & $\frac12$ & $\frac12$ & $\frac14$ & $\frac14$  & $\frac12$ & $\frac12$\\
$\rho^- \pi^0$ & $-\frac{1}{\sqrt{2}}$ & 0 & 0 & $-\frac{1}{\sqrt{2}}$ & 
$-\frac{1}{\sqrt{2}}$ & $\frac{1}{\sqrt{2}}$ & 0 & 0 & 0 & $\frac{1}{\sqrt{2}}$ & $\frac{1}{\sqrt{2}}$ & $\frac{1}{2\sqrt{2}}$ & 
$-\frac{1}{\sqrt{2}}$ & $\frac{1}{\sqrt{2}}$\\
$\pi^- \rho^0$ & 0 & $-\frac{1}{\sqrt{2}}$ & $-\frac{1}{\sqrt{2}}$ & 0 & 
$\frac{1}{\sqrt{2}}$ & $-\frac{1}{\sqrt{2}}$ & 0 & 0 & $\frac{1}{\sqrt{2}}$ & 0 & $\frac{1}{2\sqrt{2}}$ & $\frac{1}{\sqrt{2}}$ & $\frac{1}{\sqrt{2}}$ & $-\frac{1}{\sqrt{2}}$\\
\hline\hline
\end{tabular}
\end{tabular}
\caption{SU(2) relations for the decays $B \to \rho \pi$
\label{tab:rhopi}}
\begin{tabular}{c}
$\bar B \to K \bar K^*$ \\
\hline\hline
\begin{tabular}{c|c|c|c|c||cccc}
mode & $P^{(u,c,t)P}_{K\bar K^*}$ & $P^{(u,c,t)V}_{K\bar K^*}$ & 
$PA^{cP}_{K\bar K^*}$ 
& $PA^{cV}_{K\bar K^*}$ & $EW^{CP}_{K\bar K^*}$ &  $EW^{CV}_{K\bar K^*}$ 
&  $EW^{PP}_{K\bar K^*}$ &  $EW^{PV}_{K\bar K^*}$  \\
\hline
$K^{*-} K^0$ & 0 & 1 & 0 & 0 & 0 & $\frac12$ & 0 & 1\\
$\bar K^{*0} K^0$ & 0 & 1 & 1 & 1 & 0 & $\frac12$  & 0 & 1\\
$K^{*-} K^+$ & 0 & 0 & $-1$ & $-1$ & 0 & 0 & 0  & 0 \\
\hline\hline
\end{tabular}
\end{tabular}
\caption{SU(2) relations for the decays $\bar B \to K \bar K^*$
\label{tab:KKb*}}
\begin{tabular}{c}
$\bar B \to \bar K K^*$ \\
\hline\hline
\begin{tabular}{c|c|c|c|c||cccc}
mode & $P^{(u,c,t)P}_{\bar KK^*}$ & $P^{(u,c,t)V}_{\bar KK^*}$ 
& $PA^{cP}_{\bar KK^*}$ & $PA^{cV}_{\bar KK^*}$ & $EW^{CP}_{\bar KK^*}$ &  
$EW^{CV}_{\bar KK^*}$ &  $EW^{PP}_{\bar KK^*}$ &  $EW^{PV}_{\bar KK^*}$  \\
\hline
$K^{*0} K^-$ & 1 & 0 & 0 & 0 & $\frac12$ & 0 & 1 & 0\\
$K^{*0} \bar K^0$ & 1 & 0 & 1 & 1 & $\frac12$ & 0 & 1  & 0 \\
$K^{*+} K^-$ & 0 & 0 & $-1$ & $-1$ & 0 & 0 & 0  & 0 \\
\hline\hline
\end{tabular}
\end{tabular}
\caption{SU(2) relations for the decays $\bar B \to \bar K K^*$
\label{tab:KbK*}}
\begin{tabular}{c}
$\bar B \to K \rho$ \\
\hline\hline
\begin{tabular}{c|c|c|c|c|c|c|c|c||cccccc}
mode & $T^P_{K\rho}$ & $T^V_{K\rho}$ & $C^P_{K\rho}$ & 
$C^V_{K\rho}$ & $P^{(u,c,t)P}_{K\rho}$ & $P^{(u,c,t)V}_{K\rho}$ &
$PA^{cP}_{K\rho}$ & $PA^{cV}_{K\rho}$ & 
$EW^{TP}_{K\rho}$ &  $EW^{TV}_{K\rho}$ & 
$EW^{CP}_{K\rho}$ &  $EW^{CV}_{K\rho}$ & 
$EW^{PP}_{K\rho}$ &  $EW^{PV}_{K\rho}$ \\
\hline
$K^{-} \rho^+$ & 0 & $-1$ & 0 & 0 & 0 & $-1$ & 0 & 
0 & 0 & 0 & 0 & 1 & 0 & $-1$\\
$K^- \rho^0$ & $0$ & $-\frac{1}{\sqrt{2}}$ & $-\frac{1}{\sqrt{2}}$ & 0 & $0$ & $-\frac{1}{\sqrt{2}}$ & 
0 & 0 & $\frac{1}{\sqrt{2}}$ & 0 & 0 & $\frac{1}{\sqrt{2}}$ & $0$ & $-\frac{1}{\sqrt{2}}$ \\
$\bar K^0 \rho^-$ & 0 & 0 & 0 & 0 & 0 & 1 & 0 & 0 &  0 & 0 & 0 & $\frac12$  & 0 & 1 \\
$\bar K^0 \rho^0$ & 0 & 0 & $-\frac{1}{\sqrt{2}}$ & 0 & 0 & $\frac{1}{\sqrt{2}}$ &
0 & 0 & $\frac{1}{\sqrt{2}}$ & 0 & 0 & $\frac{1}{2\sqrt{2}}$ & 0 & $\frac{1}{\sqrt{2}}$\\
\hline\hline
\end{tabular}
\end{tabular}
\caption{SU(2) relations for the decays $B \to K \rho$
\label{tab:Krho}}
\begin{tabular}{c}
$\bar B \to K^* \pi$ \\
\hline\hline
\begin{tabular}{c|c|c|c|c|c|c|c|c||cccccc}
mode & $T^P_{K^*\pi}$ & $T^V_{K^*\pi}$ & $C^P_{K^*\pi}$ & 
$C^V_{K^*\pi}$ & $P^{(u,c,t)P}_{K^*\pi}$ & $P^{(u,c,t)V}_{K^*\pi}$ &
$PA^{cP}_{K^*\pi}$ & $PA^{cV}_{K^*\pi}$ &
$EW^{TP}_{K^*\pi}$ &  $EW^{TV}_{K^*\pi}$ & 
$EW^{CP}_{K^*\pi}$ &  $EW^{CV}_{K^*\pi}$ & 
$EW^{PP}_{K^*\pi}$ &  $EW^{PV}_{K^*\pi}$ \\
\hline
$K^{*-} \pi^+$ & $-1$ & $0$ & 0 & 0 & $-1$ & 0 & 0 & 0 & 0 & 0 & 1 & 0 & $-1$ & 0\\
$K^{*-} \pi^0$ & $-\frac{1}{\sqrt{2}}$ & 0 & 0 & $-\frac{1}{\sqrt{2}}$ & $-\frac{1}{\sqrt{2}}$ & 0 & 0 & 0 
& 0 & $\frac{1}{\sqrt{2}}$ & $\frac{1}{\sqrt{2}}$ & 0 & $-\frac{1}{\sqrt{2}}$ & 0\\
$\bar K^{*0} \pi^-$ & 0 & 0 & $0$ & 0 & 1 & 0 & 0 & 0 & 0 & 0 & $1/2$ & 0  & 1 & 0 \\
$\bar K^{*0} \pi^0$ & 0 & 0 & 0 & $-\frac{1}{\sqrt{2}}$ & $\frac{1}{\sqrt{2}}$ & 0 & 0 & 0 &
0 & $\frac{1}{\sqrt{2}}$ & $\frac{1}{2\sqrt{2}}$ & 0 & $\frac{1}{\sqrt{2}}$ & 0 \\
\hline\hline
\end{tabular}
\end{tabular}
\caption{SU(2) relations for the decays $B \to  K^* \pi$
\label{tab:Kspi}}
\end{table*}

The $B\to VP$ decays can be described analogously.
For this case Bose symmetry does not constrain the quantum numbers of the final
state, which allows more independent amplitudes. We define them using the convention
of \cite{B2VP} which adds one index $P$ or $V$ depending on the light meson which picks up
the spectator in the B meson. 
Our results are presented in Tables~\ref{tab:rhopi}-\ref{tab:Kspi}. The graphical 
amplitudes can again be expressed in terms of
decay constants and $B \to M$ transition matrix elements. We find 
\begin{eqnarray}\label{GPPV}
T^P_{PV} &\!\!\!=\!\!\!& -N_0 
 f_{P} \left[ \langle c_{1u} \rangle_V 
\zeta^{BP} + \langle b_{1u}\zeta_J^{BP} \rangle_{V} \right]\\
C^P_{PV} &\!\!\!=\!\!\!& -N_0 
 f_V \left[ \langle c_{2u}\rangle_V 
\zeta^{BP} + \langle b_{2u} \zeta_J^{BP} \rangle_V \right]\nn\\
P^{uP}_{PV} &\!\!\!=\!\!\!&  -N_0 f_V
\left[ \langle c_{4u}\rangle_V 
\zeta^{BP} + \langle b_{4u}  \zeta_J^{BP} \rangle_V\right]\nn\\
P^{tP}_{PV} &\!\!\!=\!\!\!&  -N_0 f_V
\left[ \langle c_{4t}^{\rm p}\rangle_V 
\zeta^{BP} + \langle b_{4t}^{\rm p} \zeta_J^{BP} \rangle_V \right]\nn\\
EW^{TP}_{PV} &\!\!\!=\!\!\!& N_0
f_V \left[ \langle c_{2t}^{\rm ew} - c_{3t}^{\rm ew} \rangle_V
\zeta^{BP} \!\!+ \langle \left(b_{2t}^{\rm ew} - b_{3t}^{\rm ew} \right)
\zeta_J^{BP} \rangle_V \right]\nn\\
EW^{CP}_{PV} &\!\!\!=\!\!\!& \frac{2N_0}{3}
f_V \left[ \langle c_{1t}^{\rm ew} \rangle_V
\zeta^{BP} + \langle b_{1t}^{\rm ew} \zeta_J^{BP} \rangle_V \right]\nn\\
EW^{PP}_{PV} &\!\!\!=\!\!\!& -\frac{N_0}{3}
f_V \left[ \langle 3\,c_{4t}^{\rm ew} + c_{1t}^{\rm ew} \rangle_V
\zeta^{BP} 
\right.\nn\\
&&\qquad\qquad\qquad \left.
+ \langle (3\, b_{4t}^{\rm ew} + b_{1t}^{\rm ew}) \zeta_J^{BP} \rangle_V \right]\,,\nn
\end{eqnarray}
where $P$ and $V$ denote any pseudoscalar or vector meson. The expressions for $G_{PV}^V$, 
where $G=\{T,C,P,EW^T,EW^C,EW^P\}$ denotes any of the graphical amplitudes, can be obtained from Eq.~(\ref{GPPV}) by taking $P \leftrightarrow V$ on the right hand side and changing the sign
between $c_{2t}^{\rm ew}$ and $c_{3t}^{\rm ew}$ in $EW^T_{PV}$.

The relations presented in Eqs.~(\ref{GPP}) and~(\ref{GPPV}) are correct to all orders 
in the perturbative matching calculation. Currently, the complete set of required 
Wilson coefficient are only 
available at tree level, and we give the resulting relations at that order in perturbation
theory. From Eq.~(\ref{cexpr}) one can see that the  Wilson coefficients $c_i(\omega_j)$ are independent of the arguments at tree level and at that order we thus find the simple relations
\begin{eqnarray}
&&\langle c_{iu} \rangle_M = c_{iu}\,,\quad
\langle c^p_{it} \rangle_M = c^p_{it}\,,\quad
\langle c^{\rm ew}_{it} \rangle_M = c^{\rm ew}_{it}\,.\qquad
\end{eqnarray}
The Wilson coefficients $b_i(\omega_j)$, given in Eq.~(\ref{bexpr}),  depend on one parameter at tree level. Thus we find 
\begin{eqnarray}
\langle b_{1u}\zeta_J^{BM_2} \rangle_{M_1} &=& \left[
      C_1 + \left(1 \!+\!\langle x^{-1} \rangle_{M_1} \right)
      \frac{C_2}{N_c}\right]\zeta_J^{BM_2}\\
 \langle b^{\rm ew}_{1t} \zeta_J^{BM_2} \rangle_{M_1}&=& 
       - \frac32 \Big[C_{10} + 
    \left(1 \!+\!\langle x^{-1}  \rangle_{M_1} \right)
      \frac{C_9}{N_c} \Big] \zeta_J^{BM_2}\nn\\
\langle b_{2u}\zeta_J^{BM_2} \rangle_{M_1} &=& \left[
      C_2 + \left(1 \!+\!\langle x^{-1}  \rangle_{M_1} \right)
      \frac{C_1}{N_c}\right]\zeta_J^{BM_2}\nn\\
 \langle b^{\rm ew}_{2t}\zeta_J^{BM_2} \rangle_{M_1} &=&  - \frac32 \Big[ C_9 + \left(1 \!+\!\langle x^{-1}  \rangle_{M_1} \right)
      \frac{C_{10}}{N_c} \Big]\zeta_J^{BM_2}
      \nn \\
 \langle b^{\rm ew}_{3t}\zeta_J^{BM_2} \rangle_{M_1} &=& 
       - \frac32  \left[ C_7 
     + \left(1 \!-\!\langle x^{-1}  \rangle_{M_1} \right)
      \frac{C_8}{N_c} \right] \zeta_J^{BM_2}
 \nn\\
 \langle b^p_{4t}\zeta_J^{BM_2} \rangle_{M_1} &=& 
  - \left[C_4 + \left(1 \!+\!\langle x^{-1}  \rangle_{M_1} \right)\frac{C_3}{N_c}\right]\zeta_J^{BM_2} 
\nn\\
 \langle b^{\rm ew}_{4t}\zeta_J^{BM_2} \rangle_{M_1} &=& 
\frac12 \left[C_{10} +(1 \!+\!\langle x^{-1}  \rangle_{M_1} )
\frac{C_9}{N_c}\right] \zeta_J^{BM_2}\,,\nn
\end{eqnarray}
and the remaining $\langle b_{i}\, \zeta_J^{BM_2} \rangle_{M_1}$ vanish at this order. Here $\langle x^{-1} \rangle_M \equiv \int_0^1 \!{\rm d}x\, x^{-1} \phi_M(x)$. 

Finally, we consider also the case of SU(3) symmetry, where  
all the amplitude parameters for $B$ decays 
to any two mesons $M_1$, $M_2$ belonging to the same SU(3) multiplets are equal
\begin{eqnarray}
G_{{P}{P}} = G \,, \,\,
G_{{V}{V}} = \widetilde{G}\,,\,\, 
G^P_{{P}{V}}= G^P\,,\,\, 
G^V_{{P}{V}} = G^V\,.
\end{eqnarray}
where again  $G=\{T,C,P,EW^T,EW^C,EW^P\}$ denotes any of the graphical amplitudes. The amplitudes with no subscripts are the 
amplitudes used in Refs.~~\cite{Chau:1990ay,Gronau:1994rj,Gronau:1998fn} and all results
presented here reproduce the well known relations of these papers. 

There is another important simplification which takes place for the EW penguin amplitudes
in limit of flavor SU(3) symmetry.  Assuming the dominance of the EWP operators
$Q_{9,10}$ one can show that 
the EWP graphical amplitudes are directly related to the tree level amplitudes~\cite{EWP,Gronau:1998fn,EWPG}. With the
normalization adopted here, these relations read
\begin{eqnarray}
& &EW(\bar K^0\pi^-) + \sqrt2 EW(K^-\pi^0) = \frac32 \kappa_+ (T+C)\nn\\
& &EW(K^-\pi^+) +  EW(\bar K^0\pi^-) = \nn\\
& &\qquad \frac34 \kappa_- (C-T)
+ \frac34 \kappa_+(C+T)\,,
\end{eqnarray}
where $EW(M_1M_2) \equiv EW^T(M_1M_2) + EW^C(M_1M_2) + EW^P(M_1M_2)$ and we have defined
\begin{eqnarray}
\kappa_\pm \equiv \frac{C_9\pm C_{10}}{C_1\pm C_2} 
\end{eqnarray}
This gives, to all orders in matching
\begin{eqnarray}
c_{1t}^{\rm ew}(\omega_j) &=& \frac34 \kappa_- [c_{1u}(\omega_j)-c_{2u}(\omega_j)] 
\nn\\
&&\qquad 
-
\frac34 \kappa_+ [c_{1u}(\omega_j) + c_{2u}(\omega_j)]\nn \\
b_{1t}^{\rm ew}(\omega_j) &=& \frac34 \kappa_- [b_{1u}(\omega_j)-b_{2u}(\omega_j)] 
\nn\\
&&\qquad 
-
\frac34 \kappa_+ [b_{1u}(\omega_j) + b_{2u}(\omega_j)] \nn\\
c_{2t}^{\rm ew}(\omega_j) &=& -\frac34 \kappa_- [c_{1u}(\omega_j)-c_{2u}(\omega_j)] 
\nn\\
&&\qquad 
-
\frac34 \kappa_+ [c_{1u}(\omega_j) + c_{2u}(\omega_j)]\nn \\
b_{2t}^{\rm ew}(\omega_j) &=& -\frac34 \kappa_- [b_{1u}(\omega_j)-b_{2u}(\omega_j)] 
\nn\\
&&\qquad 
-
\frac34 \kappa_+ [b_{1u}(\omega_j) + b_{2u}(\omega_j)] 
\label{cew2relation}
\end{eqnarray}
These relations are satisfied by the tree level Wilson coefficients in 
Eqs.~(\ref{cexpr}) and~(\ref{bexpr}), but they are correct to all orders in 
$\alpha_s(m_b)$. 

A similar relation can be written for $c_{4t}^{\rm ew}$ provided that
one makes the approximation $\kappa_+ \simeq \kappa_- \equiv \kappa$.
Numerically this is well satisfied, with $\kappa_+(m_b) = -8.75 \times 10^{-3}$ 
and $\kappa_-(m_b) = -9.31 \times 10^{-3}$ \cite{Buchalla:1995vs}. Working in this 
approximation, one has  \cite{Gronau:1998fn}
\begin{eqnarray}
EW^P_{M_1 M_2} = \kappa P^u_{M_1 M_2}
\end{eqnarray}
which gives another relation among SCET Wilson coefficients
\begin{eqnarray}
c_{4t}^{\rm ew}(\omega_i) + \frac13 c_{1t}^{\rm ew}(\omega_i) = \kappa c_{4u}(\omega_i)
\end{eqnarray}

In conclusion, the results of our paper establish the connection between the 
graphical amplitudes, often used to parameterize nonleptonic B decay
amplitudes, and the matrix elements of SCET operators. This allows an easier
interpretation of phenomenological fits to data such as those performed in 
Refs.~\cite{pheno} in terms of nonperturbative SCET parameters.
Such an analysis could help to constrain or extract the SCET parameters $\zeta^{BM}$
and $\zeta^{BM}_J$ in a way similar to the $B\to \pi\pi$ study done in 
Ref.~\cite{Bauer:2004tj}. 
Although the analysis here was limited to the leading order in the $1/m_b$ 
expansion, in principle it could be extended using the SCET expansion
to include also power corrections.
Finally, the model-independent relations 
Eqs.~(\ref{cew2relation}) fix the dominant electroweak penguin amplitudes and
provide useful checks of perturbative matching computations in SCET.

\begin{acknowledgments}
We would like to thank Iain Stewart and Ira Rothstein for collaboration on related work 
and Mark Wise for comments on the manuscript. 
This work was supported by the U.S. Department of Energy 
under Grants DF-FC02-94ER40818 and DE-FG03-92-ER-40701.
\end{acknowledgments}

\end{document}